\documentclass[preprint,aps,amsmath,amssymb]{revtex4}

\usepackage{graphicx}
\usepackage{dcolumn}
\usepackage{bm}

\begin{document}

\title{Two-loop QCD corrections to semileptonic $b$-quark decays near maximum recoil}

\author{Alexey Pak$^{1}$, Ian Blokland$^{2}$, and Andrzej Czarnecki$^{1}$}

\affiliation{
  1) Department of Physics, University of Alberta, Edmonton AB T6G 2J1, Canada \\
  2) Augustana Faculty, University of Alberta, Camrose, AB, T4V 2R3, Canada}

\date{\today}

\preprint{Alberta Thy 04-06}

\begin{abstract}
Two-gluon radiative corrections to the $b\to c\ell\bar{\nu}_\ell$ decay width
have been computed analytically as an expansion in terms of 
$\frac{m_c}{m_b}\ll 1$ in the kinematical limit of zero lepton 
invariant mass. The obtained results match smoothly with a previously known expansion
around $\left( 1 - \frac{m_c}{m_b} \right) \ll 1$. Together they describe 
the process $b\to c\ell\bar{\nu}_\ell$ for all mass ratios $\frac{m_c}{m_b}$.
\end{abstract}

\maketitle

\section{\label{sec:intro}Introduction}

Inclusive semileptonic decays of the $b$-quark can be described with a high 
degree of theoretical precision because nonperturbative effects are 
suppressed~\cite{Manohar:2000dt} and the $b$-quark mass is sufficiently large for
the perturbative QCD expansion. However, the perturbative 
calculations are challenging because of the presence of massive propagators 
of $b$- and $c$-quarks. In particular, the $\mathcal{O}(\alpha_s^2)$ 
corrections to the rate of the decay $b\to c\ell\bar{\nu}_\ell$ have not 
yet been calculated, although a rather reliable estimate was obtained in~\cite{Czarnecki:1997fc}. 
The need for such corrections has been pointed 
out in connection with a precise determination of the CKM parameter $V_{cb}$
(see, for example, \cite{Bauer:2002sh}). Such a calculation requires a generalization 
of the four-loop studies of $\mu\to e\bar{\nu}_e\nu_\mu$~\cite{vanRitbergen:1999fi} 
and $b\to u\ell\bar{\nu}_\ell$~\cite{vanRitbergen:1999gs}
to the case of a massive charged particle in the final state and is thus 
very difficult. One possible approach is to treat that final-state quark mass 
as a perturbation. In this paper we examine this approach, not with the full 
problem of the total decay rate, but in one special kinematical configuration: 
the decay $b\to c\ell\bar{\nu}_\ell$ with the lepton and neutrino escaping with 
a zero invariant mass, $q^2 = 0$. We treat the mass ratio $\frac{m_c}{m_b}$ as 
a small parameter and find several terms of expansion of 
$d\Gamma(b\to c\ell\bar{\nu}_\ell)/dq^2$ at $q^2 = 0$. This work builds on the 
recent study~\cite{Blokland:2005vq} of the decay $t\to Wb$, where an expansion 
in the $W$ mass was constructed. 

In the limit $q^2 = 0$, massless leptons are produced at rest with respect to each other 
and the corresponding decay rate may be represented as a product of $b$ decay
into a fictitious real massless $W$-boson and a consequent $W$ decay. 
The decay amplitude is 
\begin{equation}
  i\mathcal{M}(b\to X\ell\bar{\nu}_\ell) = \mathcal{J}^\mu
  \frac{-i}{q^2 - m_W^2}\left(g^{\mu\nu} - \frac{q^\mu q^\nu}{m_W^2}\right)
  \bar{u}
  \frac{ig_w}{2\sqrt{2}}\gamma^\nu\left(1 + \gamma^5\right)
  v,
\end{equation}
where $\mathcal{J}^\mu$ is the quark current.
The decay rate of $b\to X\ell\bar{\nu}_\ell$ may be represented as follows:
\begin{eqnarray}
  \nonumber
  d\Gamma(b\to X\ell\bar{\nu}_\ell) 
  &=& \frac{(2\pi)^4}{2m_b}|\mathcal{M}|^2 d\Phi(b\to X\ell\bar{\nu}_\ell) 
  \\ \nonumber
  &\stackrel{q^2\to 0}{\longrightarrow}& 
    dq^2\frac{G_F}{4\sqrt{2}\pi^2}\times
    \frac{(2\pi)^4}{2m_b}
    \mathcal{J}^\mu\mathcal{J}^{*\nu}
    \frac{q^\mu q^\nu}{m_W^2}
    d\Phi(b\to X W^*)
  \\
  &=& d\Gamma(b\to X W^*)|_{m(W^*)\to 0}\times \frac{G_F}{4\sqrt{2}\pi^2}dq^2.
\end{eqnarray}
Thus $d\Gamma(b\to c\ell\bar{\nu}_\ell)/dq^2$ at $q^2\to 0$ differs from 
$d\Gamma(b\to c W^*)$ with $m(W^*) = 0$ by only a constant factor. In this paper 
we focus on the decay width $\Gamma(b\to cW^*)$ and integrate over final 
states containing up to two gluons.

To treat virtual and real gluon emission consistently, we take advantage of 
the optical theorem which connects the $b$-quark decay rate to the imaginary 
part of the $b$-quark self-energy operator:
$\Gamma(b\to X) = \frac{1}{m_b}\mbox{Im}~{\mathcal{P}(b\to X \to b)}$.
In this approach the calculation of $\mathcal{O}(\alpha_s^2)$ corrections to 
$\Gamma(b\to c W^*)$ requires the evaluation of 3-loop diagrams instead of the 4-loop 
diagrams of the corresponding $b\to c\ell\bar{\nu}_\ell$ decay.

Several years ago~\cite{Czarnecki:1997fc}, the $b\to cW^*$ decay rate 
was found from an expansion around the so-called zero recoil limit where
$\left( 1 - \frac{m_c}{m_b} \right) \ll 1$. Our present expansion around $\frac{m_c}{m_b} \ll 1$ 
is complementary and we will find that, together with \cite{Czarnecki:1997fc}, we now know 
the differential rate at all values of the $c$-quark mass.

This paper is organized as follows. In the next section we introduce the 
notations, discuss the gauge-invariant contributions to the decay rate, 
and present some technical details of the calculation. In Section~\ref{sec:res}
we present the results and combine them with those of Ref.~\cite{Czarnecki:1997fc}
to cover the full range of final state quark mass.

\section{\label{sec:calc}$\mathcal{O}(\alpha_s^2)$ corrections to $b\to cW^*$ decay width.}

The rate of $b\to cW^*$ decay may be written as an expansion in the strong coupling 
constant $\alpha_s$ whose coefficients are functions of $\rho = \frac{m_c}{m_b}$:
\begin{eqnarray}
  \label{eq:Gamma}
  \Gamma(b\to cW^*) &=& \Gamma_0\left[X_0 + C_F\frac{\alpha_s}{\pi}X_1 
    + C_F\left(\frac{\alpha_s}{\pi}\right)^2X_2 + \mathcal{O}(\alpha_s^3)\right], \\
  \mbox{with } X_0 &=& (1 - \rho^2)^3 \\ \nonumber
  \mbox{and }
  X_1 &=& \frac{5}{4} - \frac{\pi^2}{3} 
    + \left[- \frac{11}{4} + \frac{\pi^2}{3} - 9\ln\rho\right]\rho^2
    + \left[\frac{1}{4} + \frac{\pi^2}{3} + 6\ln\rho\right]\rho^4 \\ 
    &+& \left[\frac{5}{36} - \frac{\pi^2}{3} - \frac{5}{3}\ln\rho\right]\rho^6
    + \left[\frac{65}{72} - \frac{5}{6}\ln\rho\right]\rho^8
    + \mathcal{O}(\rho^{10}) \ .
\end{eqnarray}
Here $\Gamma_0 = \frac{G_F |V_{cb}|^2 m_b^3}{8\sqrt{2}\pi}$ is the result  
corresponding to  $m_c = 0$.
The second-order correction $X_2$ may be written as a sum of finite, gauge-invariant combinations:
\begin{equation}
  \label{eq:X2}
  X_2 = T_R N_L X_L + T_R N_H X_H + T_R N_C X_C + C_F X_A + C_A X_{NA} \ .
\end{equation}
$N_L$ represents the number of massless quarks (3 in this context) while
$N_H$ and $N_C$ label the contributions of $b$- and $c$-quarks, respectively. The top quark 
contribution is suppressed by $\left(\frac{m_b}{m_t}\right)^2$ and 
we neglect it. In $SU(3)$, the color factors are 
$T_R = \frac{1}{2}$, $C_F = \frac{4}{3}$, and $C_A = 3$. 

\begin{center}
\begin{figure}[tb!]
  \includegraphics[width=0.22\textwidth]{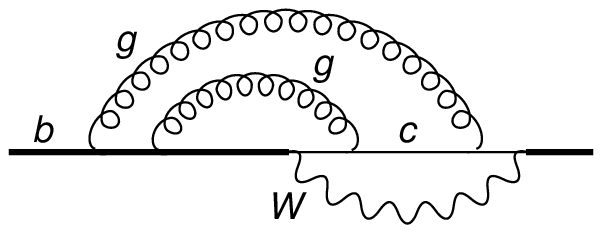}\hspace{0cm}
  \includegraphics[width=0.22\textwidth]{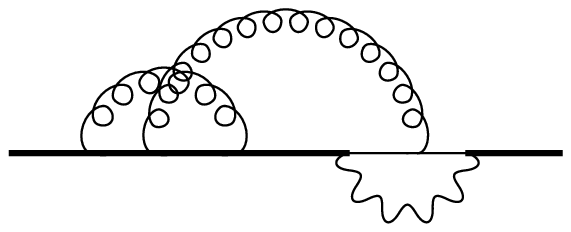}\hspace{0cm}
  \includegraphics[width=0.22\textwidth]{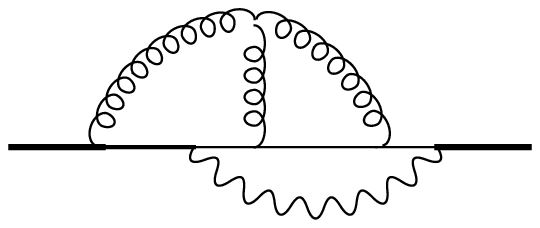}\hspace{0cm}
  \includegraphics[width=0.22\textwidth]{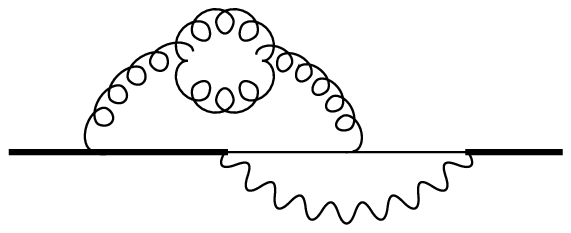}\\
  \vspace{-0.4cm}
  \parbox[t]{0.22\textwidth}{Color factor: $C_F^2$}\hspace{0cm}
  \parbox[t]{0.22\textwidth}{$C_F^2 - C_AC_F/2$}\hspace{0cm}
  \parbox[t]{0.22\textwidth}{$- C_AC_F/2$}\hspace{0cm}
  \parbox[t]{0.22\textwidth}{$C_AC_F/2$}\\
  \vspace{0.5cm}
  \includegraphics[width=0.22\textwidth]{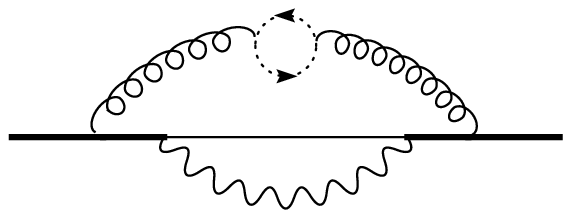}\hspace{0cm}
  \includegraphics[width=0.22\textwidth]{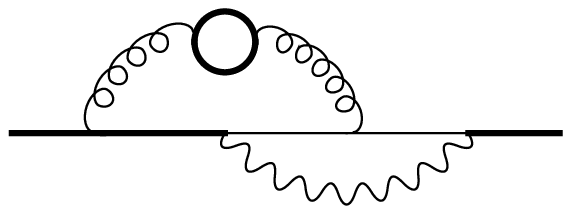}\hspace{0cm}
  \includegraphics[width=0.22\textwidth]{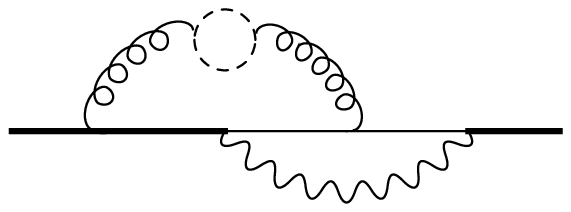}\hspace{0cm}
  \includegraphics[width=0.22\textwidth]{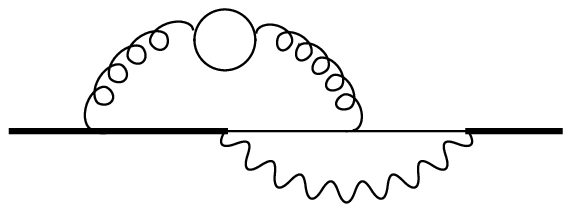}\\
  \vspace{-0.4cm}
  \parbox[t]{0.22\textwidth}{Ghost loops}\hspace{0cm}
  \parbox[t]{0.22\textwidth}{$b$-quark loops}\hspace{0cm}
  \parbox[t]{0.22\textwidth}{$u$-,$d$-,$s$-quark loops}\hspace{0cm}
  \parbox[t]{0.22\textwidth}{$c$-quark loops}\\
  \vspace{-0.2cm}
  \parbox[t]{0.22\textwidth}{($C_AC_F$)}\hspace{0cm}
  \parbox[t]{0.22\textwidth}{($C_FT_RN_H$)}\hspace{0cm}
  \parbox[t]{0.22\textwidth}{($C_FT_RN_L$)}\hspace{0cm}
  \parbox[t]{0.22\textwidth}{($C_FT_RN_C$)}\\
  \vspace{-0.3cm}
  \parbox[t]{1.0\textwidth}{\caption{\label{fig:diagrams}
    Examples of diagrams contributing to $\mathcal{O}(\alpha_s^2)$ 
    corrections to $b$-decay.}}
\end{figure}
\end{center}

\subsection{\label{subsec:calc}Calculation}

To find $X_L$, $X_H$, $X_C$, $X_A$, and $X_{NA}$ of Eq.~(\ref{eq:X2}), we 
need to consider 39 three-loop diagrams such as those in Fig.~\ref{fig:diagrams}, along 
with 19 one- and two-loop renormalization contributions.
The contributing diagrams depend in general on two scales: 
$m_b$ and $m_c$. To account for phenomena at 
different scales properly, we apply the well-known asymptotic 
methods. Table \ref{tab:regs} presents contributing momentum regions in one 
example of a three-loop double-scale topology. In each region loop 
momenta are either ``hard'' ($|k|\gg m_c$) or ``soft'' ($|k|\sim m_c$),
with $|p| = m_b$ being a hard momentum, allowing us to Taylor 
expand the propagators and thus reduce the number of scales to one. 
For example, 
\begin{equation}
  \left(|k_1|\sim m_c,~|k_3|\gg m_c\right) \ \ \ \Rightarrow \ \ \
  ~~\frac{1}{(p + k_3 - k_1)^2 + m_b^2} = 
  \sum_{n = 0}^{\infty}\frac{(2k_1p + 2k_1k_3 - k_1^2)^n}{(k_3^2 + 2k_3p)^{n + 1}} \ .
\end{equation} 
\begin{table}
\caption{\label{tab:regs}Example of the asymptotic expansion of a double-scale topology.
In the figures, dotted, thin solid, and thick solid lines represent massless, soft-scale 
massive, and hard-scale massive propagators, respectively.}
\begin{ruledtabular}
\begin{tabular}{ll}
\parbox[t]{3cm}{\vspace{-0.4cm}\includegraphics[width=2.8cm]{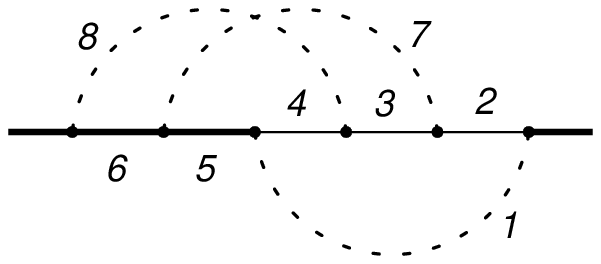}}&
\parbox[t]{13cm}{\vspace{-0.3cm}
  $I(a_1,a_2,a_3,a_4,a_5,a_6,a_7,a_8) = \int\frac{[d^Dk_1][d^Dk_2][d^Dk_3]}
    {[1]^{a_1}[2]^{a_2}[3]^{a_3}[4]^{a_4}[5]^{a_5}[6]^{a_6}[7]^{a_7}[8]^{a_8}}$, 
} \\ \multicolumn{2}{l}{\parbox[t]{16cm}{\flushleft\vspace{-0.8cm}
  $[1] = (k_1 - p)^2$, $[2] = k_1^2 + m_c^2$, $[3] = k_2^2 + m_c^2$, $[4] = k_3^2 + m_c^2$,
  $[5] = (p + k_3 - k_1)^2 + m_b^2$, \\ $[6] = (p + k_3 - k_2)^2 + m_b^2$,
  $[7] = (k_2 - k_1)^2$, $[8] = (k_3 - k_2)^2$, $(p^2 = - m_b^2)$
  \vspace{0.2cm}  }} 
\\ \hline
\parbox[t]{3cm}{\vspace{-0.3cm}\includegraphics[width=2.8cm]{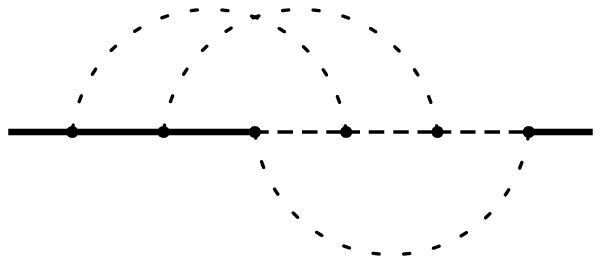}}&
\parbox[t]{13cm}{\flushleft\vspace{-0.7cm}
  Region 1: $\bf{|k_1|,|k_2|,|k_3|\gg m_c}$ (3-loop single-scale topology) \\
  $[2]\to k_1^2$, $[3]\to k_2^2$, $[4]\to k_3^2$
  \vspace{0.2cm}} 
\\ \hline
\parbox[t]{3cm}{\vspace{-0.3cm}\includegraphics[width=2.8cm]{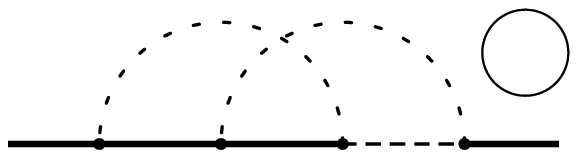}}&
\parbox[t]{13cm}{\flushleft\vspace{-0.7cm}
  Region 2: $\bf{|k_1|,|k_3|\gg m_c}$, $\bf{|k_2|\sim m_c}$ (vacuum bubble $\times$ 2-loop topology) \\
  $[2]\to k_1^2$, $[4]\to k_3^2$, $[6]\to (p + k_3)^2 + m_b^2$, $[7]\to k_1^2$,$[8]\to k_3^2$
  \vspace{0.2cm}} 
\\ \hline
\parbox[t]{3cm}{\vspace{-0.3cm}\includegraphics[width=2.8cm]{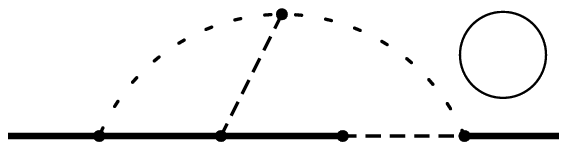}}&
\parbox[t]{13cm}{\flushleft\vspace{-0.7cm}
  Region 3: $\bf{|k_1|,|k_2|\gg m_c}$, $\bf{|k_3|\sim m_c}$ (vacuum bubble $\times$ 2-loop topology) \\
  $[2]\to k_1^2$, $[3]\to k_2^2$, $[8]\to k_2^2$, $[5]\to (p - k_1)^2 + m_b^2$,
  $[6]\to (p - k_2)^2 + m_b^2$
  \vspace{0.2cm}} 
\\ \hline
\parbox[t]{3cm}{\vspace{-0.3cm}\includegraphics[width=2.8cm]{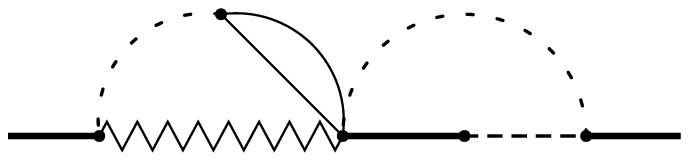}}&
\parbox[t]{13cm}{\flushleft\vspace{-0.7cm}
  Region 4: $\bf{|k_1|\gg m_c}$, $\bf{|k_2|,|k_3|\sim m_c}$ (eikonal topology $\times$ 1-loop topology) \\
  $[2]\to k_1^2$, $[5]\to (p - k_1)^2 + m_b^2$, $[6]\to 2p(k_3 - k_2)^2 + i0$, $[7]\to k_1^2$
  \vspace{0.2cm}}
\end{tabular}
\end{ruledtabular}
\end{table}

Table~\ref{tab:regs} illustrates the asymptotic expansion process for one 
of the integrals encountered in this work. In this example, Region 4 
involves so-called eikonal integrals, featuring the propagator $2p(k_3 - k_2) + i0$.
Although seemingly double-scale, such integrals only multiplicatively 
depend on the external momentum $p$. Care should be taken with eikonal 
regions, since the integral value may depend on 
the sign of the contour fixing term $i0$.

The large number of resulting integrals can be reduced to a small set of 
``master integrals'' (see Ref. \cite{Laporta:2001dd} for an example of 
a solution algorithm along with references to earlier work). 
\begin{figure}[bt]
  \vspace{5mm}
  \includegraphics[width=0.4\textwidth]{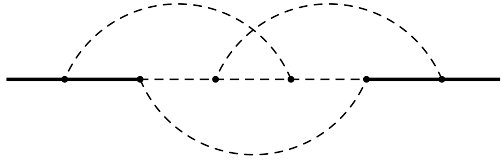}
  \caption{\label{fig:t3i}The new master integral of Eq.~(\ref{eq:mif}). 
    The solid and dashed lines correspond to massive and massless propagators, respectively.}
\end{figure}
Most of the master integrals used in this paper have been 
described in~\cite{Blokland:2005vq}. An additional master 
integral corresponds to the topology in Fig.~\ref{fig:t3i}; for completeness 
we present here the result using the same notations:
\begin{eqnarray}
  \nonumber
  \mbox{Im}~F(1,1,1,1,1,1,1,1,0) &=& 
  \mbox{Im}~\int\frac{[d^Dk_1][d^Dk_2][d^Dk_3]}
  {k_1^2 k_2^2 k_3^2 (k_2^2 + 2k_2p) (k_3^2 + 2k_3p)} 
  \\ \nonumber &\times& 
  \frac{1}{(p + k_1 + k_2)^2 (p + k_1 + k_3)^2 
    (p + k_1 + k_2 + k_3)^2} \\
  &=& \pi\mathcal{F}^3\left[\frac{61\pi^4}{360} 
    + \mathcal{O}(\epsilon)\right],
  \label{eq:mif}
\end{eqnarray}
where $\mathcal{F} = \frac{\Gamma(1 + \epsilon)}{(4\pi)^{D/2}}$ is a 
common loop factor and $D = 4 - 2\epsilon$ is the convention used for dimensional regularization .

\section{\label{sec:res}Results}

\begin{figure}[hbt!]
  \includegraphics[width=0.48\textwidth]{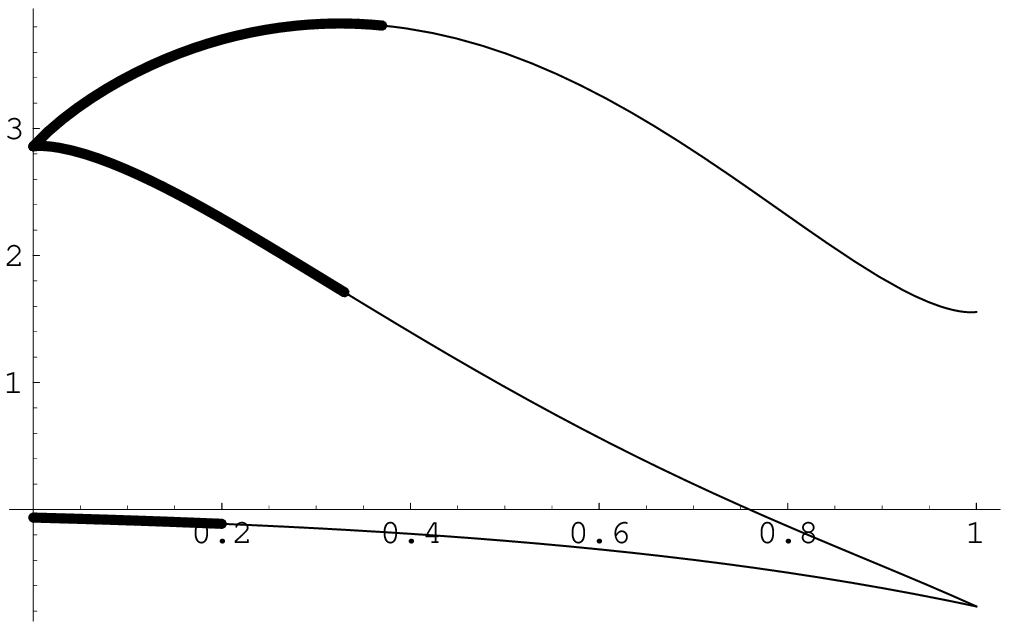}
  \put(-15,35){\makebox(0,0){$\rho$}}  
  \put(-155,120){\makebox(0,0){$X_L / (1-\rho)^3$}}
  \put(-95,70){\makebox(0,0){$X_C / (1-\rho)^3$}}
  \put(-165,10){\makebox(0,0){$X_H / (1-\rho)^3$}}
  \hspace{0.2cm}  
  \includegraphics[width=0.45\textwidth]{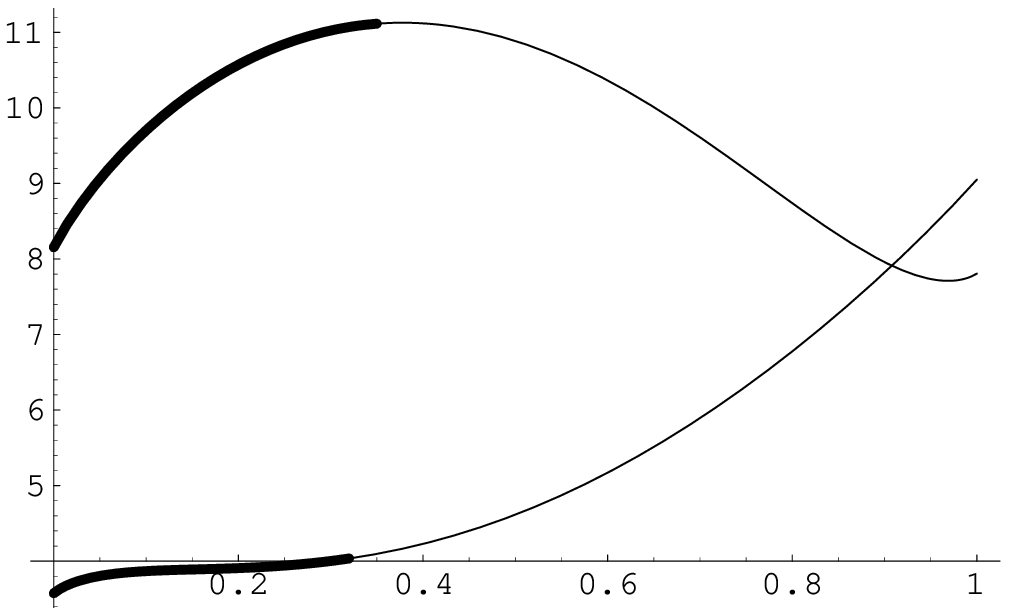}
  \put(-13,20){\makebox(0,0){$\rho$}}
  \put(-155,25){\makebox(0,0){$X_A / (1-\rho)^3$}}
  \put(-130,100){\makebox(0,0){$- X_{NA} / (1-\rho)^3$}}
  \\ \vspace{-0.5cm}
  \parbox[t]{1.0\textwidth}{\caption{\label{fig:results}
    Matching of expansions around $\rho = 0$ (thick line) and $\rho = 1$ (thin line).
    The $c$-quark contribution $X_C$ interpolates between heavy ($X_H$) and light 
    ($X_L$) quark contributions.}}
\end{figure}

Our results for the contributions to Eq.~(\ref{eq:X2}) are obtained as series 
in $\rho=\frac{m_c}{m_b}$. We have obtained terms up to $\rho^9$ for the expansions 
and present here contributions to $X_2$ through terms of order $\rho^3$:
\begin{eqnarray}
  X_L &=& - \frac{4}{9} + \frac{23\pi^2}{108} + \zeta_3 
     + \left[\frac{28}{9} - \frac{101\pi^2}{108} - \zeta_3
       + \frac{13}{2}\ln\rho - 3\ln^2\rho\right]\rho^2
     + \mathcal{O}(\rho^4),\\ 
  X_H &=& \frac{12991}{1296} - \frac{53\pi^2}{54} - \frac{\zeta_3}{3}
     + \left[ - \frac{135467}{6480} + \frac{113\pi^2}{54} 
       + \frac{\zeta_3}{3}\right]\rho^2
     + \mathcal{O}(\rho^4),\\ 
  \nonumber
  X_C &=&  - \frac{4}{9}  + \frac{23\pi^2}{108} + \zeta_3 - \frac{3\pi^2}{4}\rho
     + \left[\frac{65}{18} + \frac{133\pi^2}{108} - \zeta_3 
  \right. \\ &+& \left. 
       \frac{13}{2}\ln\rho - 3\ln^2\rho\right]\rho^2 - \frac{25\pi^2}{18}\rho^3
     + \mathcal{O}(\rho^4),\\ 
  \nonumber 
  X_A &=& 5 - \frac{119\pi^2}{48} - \frac{53}{8}\zeta_3
     - \frac{11\pi^4}{720} + \frac{19\pi^2}{4}\ln{2} 
     + \left[- \frac{315}{8} + \frac{497\pi^2}{48} + \frac{151}{8}\zeta_3 
  \right. \\ &+& \left. 
       \frac{37\pi^4}{360} - \frac{57\pi^2}{4}\ln{2}
     + \left(\pi^2 - \frac{75}{8}\right)\ln\rho 
       - \frac{27}{2}\ln^2\rho\right]\rho^2 - \frac{4\pi^2}{3}\rho^3
     + \mathcal{O}(\rho^4),\\ 
  \nonumber 
  X_{NA} &=& \frac{521}{576} + \frac{505\pi^2}{864} + \frac{9}{16}\zeta_3
     + \frac{11\pi^4}{1440} - \frac{19\pi^2}{8}\ln{2}
     + \left[- \frac{2315}{576} - \frac{2119\pi^2}{864}
  \right. \\  &-& \left.
       \frac{107}{16}\zeta_3 - \frac{\pi^4}{144} + \frac{57\pi^2}{8}\ln{2}
       - \frac{185}{8}\ln\rho + \frac{33}{4}\ln^2\rho\right]\rho^2
       + \frac{2\pi^2}{3}\rho^3
     + \mathcal{O}(\rho^4). 
\end{eqnarray}
We have used the $\overline{\rm MS}$ definition of $\alpha_s$ normalized at $\mu = m_b$, 
and the pole mass $m_b$.

These expansions may be directly compared to expansions around $\rho = 1$ 
of~\cite{Czarnecki:1997fc} as follows:
\begin{eqnarray}
  \label{eq:xd1}
  X_L(\rho) &\to& (1 - \rho)^3 \Delta_L(1 - \rho), \\
  \label{eq:xd2}
  X_H(\rho) &\to& (1 - \rho)^3 \left[\Delta_H(1 - \rho) - \Delta_C(1 - \rho)\right], \\
  \label{eq:xd3}
  X_C(\rho) &\to& (1 - \rho)^3 \Delta_C(1 - \rho), \\
  \label{eq:xd4}
  X_A(\rho) &\to& (1 - \rho)^3 \Delta_F(1 - \rho), \\
  \label{eq:xd5}
  X_{NA}(\rho) &\to& (1 - \rho)^3 \left[\Delta_A(1 - \rho) 
    - \frac{1}{2}\Delta_F(1 - \rho)\right].
\end{eqnarray}

In \cite{Czarnecki:1997fc} corrections from $b$- and $c$-quarks were lumped together
in $\Delta_H$. Here we divide them up by separating the $c$ contribution in 
$\Delta_C$ (Eq.~(\ref{eq:deltac})). As expected, $X_C$ reaches $X_L$ in the 
limit $\rho\to 0$ and $X_H$ in the limit $\rho\to 1$.

Plots on Fig.~\ref{fig:results} present $X_L$, $X_H$, $X_C$,
$X_A$, and $X_{NA}$ calculated to $\mathcal{O}(\rho^{10})$ 
around $\rho = 0$, and expansions of the corresponding functions from 
\cite{Czarnecki:1997fc} (Eq. (A1) of that work), according to Eqs.~(\ref{eq:xd1})-(\ref{eq:xd5}),
calculated through $\mathcal{O}((1 - \rho)^{21})$. 
It is sufficient to account for terms up to $\mathcal{O}(\rho^7)$ to reach the 
relative accuracy of 1\% at the realistic value of $m_c/m_b\approx 1/3$.

For convenience we provide results for a numerical fit providing accuracy better 
than $0.01$ for $X_L$, $X_C$, $X_A$, and $X_{NA}$, and better than $10^{-5}$ for 
$X_H$ for $0\le\rho\le 1$:
\begin{eqnarray}
  X_L/(1 - \rho)^3 &\approx& 2.872 + 6.849\rho - 17.00\rho^2 
                           + 22.56\rho^3 - 26.92\rho^4 + 13.16\rho^6, \\
  X_H/(1 - \rho)^3 &\approx& -0.06361 - 0.1902\rho - 0.2378\rho^2 
                           - 0.1733\rho^3 - 0.09828\rho^4, \\
  X_C/(1 - \rho)^3 &\approx& 2.882 - 0.9432\rho - 14.31\rho^2 + 25.00\rho^3 
                           - 18.49\rho^4 + 5.113\rho^5, \\ \nonumber
  X_A/(1 - \rho)^3 &\approx& 3.531 + 1.305\sqrt{\rho} + 0.1496\rho - 13.76\rho^2 
       + 49.64\rho^3 - 57.77\rho^4 \\ &+& 33.69\rho^5 - 7.746\rho^6, \\ \nonumber
  X_{NA}/(1 - \rho)^3 &\approx& -8.090 - 1.696\sqrt{\rho} - 12.77\rho 
       + 20.35\rho^2 + 8.257\rho^3 - 43.23\rho^4 \\ &+& 58.52\rho^5 - 29.09\rho^6.
\end{eqnarray}

\section{\label{sec:conclusion}Conclusion}

\begin{figure}[hbt!]
  \includegraphics[width=0.4\textwidth]{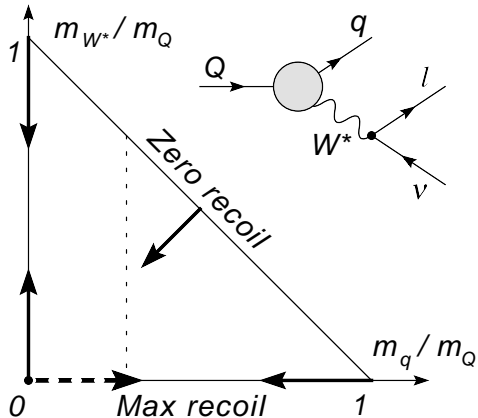}
  \caption{\label{fig:triangle}Kinematic boundaries of semileptonic heavy quark decays. 
  The dotted arrow shows the expansion presented in this paper. Previously 
  known expansions are indicated with solid arrows. Integration over the dotted 
  line would correspond to integration over the leptonic phase space in the decay 
  $b\to c\ell\bar{\nu}_\ell$.}
\end{figure}

Fig.~\ref{fig:triangle} illustrates existing expansions \cite{Czarnecki:1997fc,Blokland:2005vq,Czarnecki:2001cz} 
of semileptonic quark decays in various kinematic configurations. Analytic expressions are known 
along the zero recoil line and in all corners of the triangle. With the 
approach demonstrated in this paper and with improving  
computational resources, it is becoming feasible to calculate the complete 
$\mathcal{O}(\alpha_s^2)$ correction to the decay rate $\Gamma(b\to c\ell\bar{\nu})$
as an expansion around $\rho = 0$. Like in the present study, the most challenging 
hard-scale regions will contribute only to even powers of $\rho$, with odd 
powers originating from factorized regions. Assuming this series will converge 
similarly to the expansion presented here, a 5\% accuracy of the complete 
correction at realistic value $\rho = \frac{1}{3}$ will require calculating terms 
through $\rho^4$ of the most difficult diagrams. An extension 
to $\rho^5$ will require evaluating only factorized diagrams and will likely 
to improve the accuracy to the level of 3\%. This is a challenging but feasible task.

In the future, if the need arises, these expansion techniques could be 
applied for computing precision rates of other heavy colored particle 
decays, e.g. squarks, accounting for mass-dependent effects. 

\emph{Acknowledgements:} We would like to thank Maciej {\' S}lusarczyk for 
helpful discussions of the Laporta algorithm and continuing support of its implementation.
Some of our algebraic calculations were done using FORM~\cite{Vermaseren:2000nd}.

\bibliography{paper}

\appendix

\section{\label{subsec:deltac}Soft quark contributions}

The contribution of $c$-quark loops expanded in $\delta = 1 - \frac{m_c}{m_b}$
was calculated as part of the study in Ref. \cite{Czarnecki:1997fc} but not explicitly 
shown. Here, for completeness, we present that result through terms $\mathcal{O}(\delta^8)$:
\begin{eqnarray}
\label{eq:deltac}
 \nonumber
 \Delta_C &=& \frac{230}{9} - \frac{8\pi^2}{3} 
   + \left[- 69 + \frac{22\pi^2}{3}\right]\delta
   + \left[\frac{15005}{162} - \frac{262\pi^2}{27}\right]\delta^2 
   \\ \nonumber
   &+& \left[- \frac{91051}{1620} + \frac{695\pi^2}{108} 
   - \frac{32}{9}\ln{2\delta}\right]\delta^3 
   + \left[\frac{1517}{405} - \frac{77\pi^2}{135}\right]\delta^4
   \\ \nonumber
   &+& \left[\frac{1002319}{56700} - \frac{3751\pi^2}{2160} - \frac{88}{135}\ln{2\delta}\right]\delta^5 
   + \left[- \frac{60481}{7560} + \frac{13033\pi^2}{15120} - \frac{88}{135}\ln{2\delta}\right]\delta^6
   \\
   &+& \left[\frac{2773441}{1587600} - \frac{493\pi^2}{3780} - \frac{586}{945}\ln{2\delta}\right]\delta^7 
   + \left[\frac{140572}{297675} - \frac{2\pi^2}{405} - \frac{556}{945}\ln{2\delta}\right]\delta^8.
\end{eqnarray}

\end{document}